\documentclass[prl,aps,twocolumn,amssymb,amsmath]{revtex4}
\usepackage{graphicx}
\newcommand{\figwidth}{.9\columnwidth}

\begin{document}
\title{X-ray diffraction of a disordered  charge density wave}
\author{Alberto Rosso}
\author{Thierry Giamarchi}
\affiliation{Universit\'e de Gene\`eve,
DPMC, 24 Quai Ernest Ansermet, CH-1211 Gen\`eve 4, Switzerland}

\begin{abstract}
We study the X-ray diffraction spectrum produced by a
collectively pinned charge density wave (CDW), for which one can
expect a Bragg glass phase. The spectrum consists of two
asymmetric divergent peaks. We compute the shape of the peaks, and
discuss the experimental consequences.
\end{abstract}
\maketitle

The statics and dynamics of  disordered elastic objects govern the
physics of a wide range of systems, either periodic, such as
vortex flux lines \cite{blatter_vortex_review} and charge density
waves (CDW) \cite{gruner_revue_cdw}, or involving propagating
interfaces, such as domain walls in magnetic
\cite{lemerle_domainwall_creep} or ferroelectric
\cite{tybell_ferro_creep} systems, contact lines of liquid menisci
on rough substrates \cite{moulinet_contact_line} and propagation
of cracks in solids \cite{gao_crack_elasticity}. It was recently
shown that periodic systems have unique properties, quite
different from the ones of the interfaces. If topological defects
(i.e. dislocations etc.) in the crystal are excluded,
displacements grow only logarithmically
\cite{nattermann_pinning,korshunov_variational_short,giamarchi_vortex_short},
instead of the power-law growth as for interfaces. The positional
order is only algebraically destroyed
\cite{giamarchi_vortex_short,giamarchi_vortex_long} leading to
divergent Bragg peaks and a nearly perfect crystal state. Quite
remarkably, it was shown that for weak disorder this solution is
\emph{stable} to the proliferation of topological defects, and
thus that a thermodynamically stable phase having both glassy
properties and quasi-long range positional order exists
\cite{giamarchi_vortex_long}. This phase, nicknamed Bragg glass,
has prompted many further analytical and experimental studies (see
e.g. \cite{nattermann_vortex_review,giamarchi_vortex_review} for
reviews and further references). Although its existence 
 can be tested indirectly by the consequences on the phase
diagram of vortex flux lines, the most direct proof is to measure
the predicted algebraic decay of the positional order. Such a
measurement can be done by means of diffraction experiments, using
either neutrons or X-rays on the crystal. Neutron diffraction
experiments have recently provided unambiguous evidence
\cite{klein_brglass_nature} of the existence of the Bragg glass
phase for vortex lattices.

Another periodic system in which one can expect a Bragg glass to
occur are charge density waves \cite{gruner_revue_cdw}, 
where the electronic density is spatially modulated. Disorder
leads to the pinning of the CDW \cite{fukuyama_pinning}. In such
systems very high resolution X-rays experiments can be performed
\cite{rouzieres_structue_cdw}. The resolution is in principle much
higher than the one that can be achieved by neutrons for vortex lattices,
consequently CDW systems should be prime candidates to check for the
existence of a Bragg glass state. However, compared to the case of
vortex lattices the interpretation of the spectrum is much
more complicated for two main reasons: (i) the phase of the CDW is
the object described by an elastic energy, whereas the
X-rays probe the displacements of the atoms in the crystal lattice
(essentially a cosine of the phase); (ii) since the impurities
substitute some atoms of the crystal, the very presence of the
impurities changes the X-ray spectrum. This generates non-trivial
terms of interference between disorder and atomic displacements
\cite{ravy_x-ray_peakasymmetry,rouzieres_structue_cdw}. It is thus
necessary to make a detailed theoretical analysis for the
diffraction due to a pinned CDW. The study of the spectrum has been
carried out so far either for strong pinning or at high
temperatures
\cite{rouzieres_structue_cdw,ravy_x-ray_whiteline,ravy_x-ray_peakasymmetry,
brazovskii_x-ray_cdwT}.

In this paper we focus on the low temperature limit where a well
formed CDW exists and on weak disorder, for which one expects to
be in the Bragg glass regime. We show that the diffraction
spectrum consists in two asymmetric peaks. In contrast to previous
assumptions \cite{ravy_x-ray_peakasymmetry},
we show that the asymmetry is present also in the
weak pinning limit. The peaks are power-law divergent, with an
anisotropy in shape. This form is consistent with the Bragg glass
behavior \cite{giamarchi_vortex_long}. The asymmetry is a
subdominant power-law too, with an exponent that we determine. We also
briefly discuss the role of unscreened Coulomb interaction for the
CDW on the diffraction spectrum.

The general expression \cite{guiner_xray}
for the total diffraction intensity in a crystal is given by
\begin{eqnarray}
    \label{eq:Sdef}
    I(q) = \frac{1}{V} \sum_{i,j} e^{-iq(R_i-R_j)} \left\langle
\overline{f_if_j e^{-iq(u_i-u_j)}} \right\rangle,
\end{eqnarray}
where  $u_i$ is the atom displacement from the equilibrium
 position $R_j = ja$, with $a$ indicating the lattice constant, $f_i$ the
 atomic scattering factor and $\overline{\langle\ldots \rangle}$
 denotes the double average over the disorder and over the thermal
 fluctuations.
As an example let us first consider  the case of fixed atoms
($u_i=0$). We obtain:
\begin{eqnarray}
\label{Laue}
I(q) = \overline{f}^2  \sum_K \delta(q-K)
+  {\Delta f}^2 N_I,
\end{eqnarray}
where $\Delta f= f_I -f $ is the difference between the impurity $I$ and the
 host
 atom scattering factors, $N_I=n_I(1-n_I) $, with $n_I$ is
the impurity concentration and
 $\overline{f}$ is the average scattering factor.
The  usual Bragg peaks, in correspondence to the  reciprocal lattice vectors
 $K$,
arise from the first term in (\ref{Laue}), the second term is
responsible for a background intensity,  called Laue
 scattering, due to the disorder.

In a second stage we  take into account  displacements of the atoms
 related to the presence of a CDW.
To this purpose, we consider an electron density characterized by a
sinusoidal modulation:
 \begin{equation}
 \rho(x) = \rho_0 \cos(Qx+\phi(x)).
\label{rho}
\end{equation}
$\phi$ is the phase of the charge density wave and $Q=2k_F$, where $k_F$
 is the Fermi wave vector.
The  associated  Hamiltonian writes:
\begin{eqnarray}
\label{eq:Hamiltonian}
H= \int d^dx \frac{c}{2}  \left( \nabla \phi(x) \right)^2
  \pm V_0\int d^dx \Sigma(x) \rho(x),
\end{eqnarray}
where $d$ is the dimension of the space. The first term in the
Hamiltonian (\ref{eq:Hamiltonian}) represents the elasticity. The
elasticity is in fact anisotropic \cite{feinberg_cdw} along the
$Q$-direction:
\begin{eqnarray} \label{eq:anisotropy}
 H_{\text{el.}}= \int d x d^{d-1}y
 \frac{c_1}{2}
 \left( \partial_x \phi \right)^2 +
 \frac{c_2}{2}  \left( \partial_y \phi \right)^2 ,
\end{eqnarray}
where $x \parallel Q$, and $c_1 \gg c_2$. The compression along
$x$ corresponds to an increase of electric charge density and thus
pays the price of Coulomb repulsion, while distortions along the
remaining $d-1$ directions are much easier.  We are led back  to
(\ref{eq:Hamiltonian}) by redefining the spatial variables $x' =
x/\sqrt{c_1}$ and  $y' = y/\sqrt{c_2}$, with
$c=(c_1c^{d-1}_2)^{\frac{1}{2}}$. The main effect in the
diffraction spectrum is thus to make the \emph{shape} of the peaks
anisotropic, but this will not change the overall divergence. The
local, but anisotropic, elasticity (\ref{eq:anisotropy}) is valid
beyond the distance at which the Coulomb interaction between
various parts of the CDW is screened. If this length is very
large, or if one want to examine short range regime one should
keep the $q$-dependence in the elastic constants. This leads to a
more complicated behavior that we will only briefly discuss here
and will be examined in details elsewhere \cite{rosso_cdw_long}.
The second term in the Hamiltonian (\ref{eq:Hamiltonian}) reflects
the effect of the disorder on the electron density. The Gaussian
random function $\Sigma(x)$
 describes the impurity  distribution and is characterized by the correlator
 $\overline{\Sigma(x)\Sigma(y)} = N_I \delta(x-y) $,
$V_0$ is a positive constant which measures the impurity potential and  finally
 the sign $+$ ($-$) is related to the  repulsive (attractive)
 interaction between the electrons and the local impurity.
 In the following, we restrict our analysis to the repulsive case, $\rho_0$ is
 absorbed in $V_0$ and we define  the disorder strength $D= {V_0}^2 N_I$.

A density modulation is accompanied by a lattice distortion $u$
given at low temperature by
\begin{equation}
 u(x) = \frac{u_0}{Q} \nabla \cos(Qx+\phi(x)).
\end{equation}
We are interested in the behavior of the scattering intensity
$I(q)$ near a Bragg peak ($q \sim K$). Since  $|\delta q| =|q-K|
\ll K$,  we can take the continuum limit  $i \rightarrow x$ and we
obtain from (\ref{eq:Sdef}):
\begin{eqnarray}
    \label{eq:Sdef2}
    I(q) =  \int_r \left\langle
\overline{f_{\frac{r}{2}}f_{-\frac{r}{2}}
 e^{-i \delta q (u(\frac{r}{2})-u(-\frac{r}{2}))}} \right\rangle.
\end{eqnarray}
where   $\int_r = \frac{1}{a^d} \int d^dr e^{-i \delta q r}$ and
$f_{\frac{r}{2}} = \overline{f}+ \Delta f a^{d/2}
\Sigma(\frac{r}{2})$. In  (\ref{eq:Sdef2}) we have applied the
standard decomposition in
 center of
mass $R$ and relative  $r$ coordinates ($x = R+\frac{r}{2}$ and
 $y = R-\frac{r}{2}$). The integration
over $R$ has already  been performed because $u$ vary slowly at
the scale of the
lattice spacing.
Assuming  that in the elastic approximation
displacements remain small ($u_i \ll R_i$),
 one can expand
(\ref{eq:Sdef2}) as powers of $Ku_0$.
 Developing up to the second order  we get \cite{ravy_x-ray_whiteline} :
\begin{eqnarray}
\label{eq:Intensities}
& &  I(q) = I_{\text{d}} + I_{\text{a}}+ I_{\text{tripl.}}, \mbox{with} \\
& & I_{\text{d}} =\overline{f}^2 q^2  \int_r
 \langle \overline{u(\frac{r}{2}) u(-\frac{r}{2})} \rangle, \nonumber \\
& & I_{\text{a}} =  -iq {\Delta f} a^{d/2}\overline{f} \int_r   \left\langle
\overline{\Sigma(-\frac{r}{2}) u(\frac{r}{2}) - \Sigma(\frac{r}{2})
 u(-\frac{r}{2}) } \right \rangle, \nonumber \\
& & I_{\text{tripl.}}= -
iq {\Delta f}^2 a^{d} \int_r
 \left\langle
\overline{\Sigma(-\frac{r}{2})\Sigma(\frac{r}{2}) (u(\frac{r}{2})
- u(-\frac{r}{2})) } \right \rangle . \nonumber
\end{eqnarray}
 While the contribution $I_{\text{d}}$ represents the intensity due  to the atomic
 displacements alone,
 the contributions $I_{\text{a}}$ and $I_{\text{tripl.}}$ are generated by the coupling
 between
the disorder and the displacement.
The presence of a CDW is signaled by the formation, around
each   Bragg peak,  of two satellites at reciprocal vectors $K \pm Q$.
In absence of disorder  ($D=0$ and $\phi \sim \mbox{const.}$) the
 displacement term has the form
 $I_{\text{d}}=f^2 q^2 u_0^2 \sum_K \delta(q+K\pm Q)$ and  the other
 terms are vanishing:
in this case the two satellites have the same intensity
and the broadening is absent.
To interpret the experimental findings
 \cite{ravy_x-ray_peakasymmetry,brazovskii_x-ray_cdwT,rouzieres_structue_cdw},
in particular to explain the measured  strong asymmetry
\footnote{ The asymmetry of the two peaks can be very strong:
 it has been observed that the intensity
of the lowest peak can even be smaller than the Laue scattering
intensity. In this case one talks of white line in the spectrum.}
between the peaks at $K+Q$ and at  $K-Q$, we need to account for
the effect of  impurities.

\begin{figure}
\centerline{ \includegraphics[width=\figwidth]{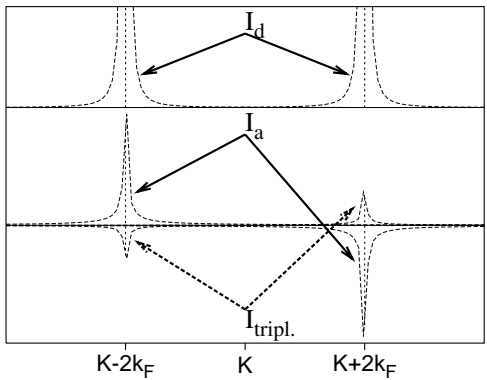} }
\caption{Intensities of the different contributions to satellite
peaks. The more divergent term, $I_{\text{d}}$, is symmetric.
 $I_{\text{a}}$ and $I_{\text{tripl.}}$  are antisymmetric, with $I_{\text{a}} \gg I_{\text{tripl.}}$.}
\label{model}
\end{figure}

In the literature  the term $I_{\text{a}}$ was evaluated  by means of
models
\cite{ravy_x-ray_whiteline,ravy_x-ray_peakasymmetry,brazovskii_x-ray_cdwT}
which describe the pinning by imposing a constant value $\phi_0$
to the phase in (\ref{rho}) in proximity of each impurity, and
$I_{\text{tripl.}}$ was conjectured  to be negligible
\cite{ravy_x-ray_whiteline,cowley_x-ray_cdw}. In that approach,
the observed satellite asymmetry is seen as a clear sign of the
strong disorder; in fact, $\phi_0$ is not  constant and, for
sufficiently large domains, one should have
\cite{ravy_x-ray_peakasymmetry} $I_{\text{a}} \propto
\overline{\cos(\phi_0)} \sim 0$. To go beyond this
phenomenological approach and also deal with the weak disorder
limit, in which one expects the Bragg glass, we use a Gaussian
variational approach
\cite{mezard_variational_replica,giamarchi_vortex_long}. We first
perform the average over the  disorder using the standard replica
techniques. The replicated Hamiltonian corresponding to
(\ref{eq:Hamiltonian}) is
\begin{equation} \label{eq:ham}
 H_{\text{\text{eff.}}}= \int d^dr\sum_a \frac{c}{2}
 (\nabla \phi_a)^2 - \frac{D}{T} \sum_{a,b}
 \cos\left(\phi_a(r) -\phi_b(r)\right) ,
\end{equation}
where $T$ is the temperature and the sum over the $n$ replica has
to be considered in the limit $n \rightarrow 0$. We stress  that,
moving from the Hamiltonian (\ref{eq:Hamiltonian}) to its
replicated version we also need to change  the correlation
functions  containing explicitly the disorder: we have, for
example,  $\left\langle \overline{\Sigma(-\frac{r}{2})
u(\frac{r}{2}) } \right \rangle \rightarrow -\frac{D}{TV_0}
\sum_{a} \left\langle \rho_a(-\frac{r}{2}) u_1(\frac{r}{2})
  \right \rangle_{\text{eff.}}$.
After some manipulations and using (\ref{rho}), we obtain:
\begin{eqnarray} \label{eq:Correlations}
&& I_{\text{d}}  =  \overline{f}^2 q^2 {u^2}_0 \int_r\left[ e^{-i Q
 r} +c.c.\right] C_{\text{d}}(r) \,,
   \\
 &&I_{\text{a}}  = - \overline{f}  {\Delta f} q  u_0 \sqrt{N_I a^d D }
\int_r \left[ e^{-i Q r} - c.c.\right]
 C_{\text{a}}(r)  \nonumber \,,
\end{eqnarray}
where $C_{\text{d}}(r)=
\left\langle  e^{i(\phi_1(\frac{r}{2})
-\phi_1(-\frac{r}{2}))}\right\rangle_{\text{eff.}}$ and
 $C_{\text{a}}(r) =\frac{1}{Tn} \sum_{a,b}^n
\left\langle  e^{i(\phi_a(\frac{r}{2}) -\phi_b(-\frac{r}{2}))}
\right \rangle_{\text{eff.}}$ are the positional correlation
functions controlling the behavior of each contribution. We notice
that the  intensity of   the peaks at $q=Q+K$ and $q=K-Q$ is
symmetric, as in the case of a pure system, for the displacement
term $I_{\text{d}}$, but it is antisymmetric for $I_{\text{a}}$. The sum of these
two terms leads to an asymmetry of the peaks. Fig.~\ref{model}
show the behavior of the different contributions.

Following the method used in \cite{giamarchi_vortex_long} for flux
lines in presence of weak disorder, we can calculate the various
terms in (\ref{eq:Correlations}). We look for the best trial
Gaussian Hamiltonian $H_0= \int_q G_{ab}(q) \phi_a(q) \phi_b(-q)$
in replica space, which approximates (\ref{eq:ham}). Defining
\begin{eqnarray}
 B_{ab}(r)&=& \langle (\phi_a(r) -\phi_b(0))^2 \rangle_0 \\
          &=&2 T\int_q \left[\tilde{G}(q) -G_{ab}(q)
 \cos qr \right], \nonumber
\end{eqnarray}
where $\tilde{G}$ is the diagonal element of $G_{ab}$, and using
the Gaussian approximation, the positional correlation functions
become  $C_{\text{a}}(r)
=    \frac{1}{nT} \sum_{a,b}^n e^{-  \frac{ B_{ab}(r)}{2}}
$ and $ C_{\text{d}}(r) = e^{-\frac{\tilde{B}(r)}{2}}$,
where $\tilde{B}$ is the diagonal element of $B_{ab}$.
 Two general classes of solutions exist for this problem: while
the first class preserves the permutation symmetry of the replica
(RS), the second class (RSB) breaks the  replica symmetry. It has
been  shown \cite{giamarchi_vortex_long} that the stable solution
for $d>2$ corresponds to the RSB class, while the RS solution
remains valid at short distance. $C_{\text{d}}$ is similar to the
correlation calculated for flux lines \cite{giamarchi_vortex_long}
and will be discussed later. To evaluate the contribution of the
interference between disorder and displacement we factorize the
antisymmetric term $C_{\text{a}}(r)=\chi(r)C_{\text{d}}(r)$. We first consider the
RS approximation:
\begin{eqnarray}
 \chi(r)=\frac{1}{T}\left[1 - e^{-T  \int_q G_c(q)\cos qr} \right],
\end{eqnarray}
where $G_{c}=\frac{1}{cq^2}$ is the connected part of $G_{ab}$.
In $d=3$  we  estimate
\begin{eqnarray}
   C^{RS}_{a}(r)
             &\sim&    \frac{2 \pi^2 }{ c r}
 e^{-  \frac{ \tilde{B}(r)}{2}}.
\end{eqnarray}
The triplet term can be evaluated in an analogous way, but it gives
 non-zero
contributions only considering higher order harmonic terms in the
electron density. Equation (\ref{rho}) becomes: $\rho(x) = \rho_0
\cos(n(Qx+\phi(x)))$, with $n=1,2$. As we have already found  for
$I_{\text{a}}$, we get an antisymmetric term
 with a prefactor
 $\propto {\Delta f}^2 q   u_0 N_i a^d D
  $ and a correlation $C_{\text{tripl.}}=\frac{1}{2nT^2} \sum_{a,b,c}^n \langle(e^{- i (\phi_c(\frac{r}{2})  -2
  \phi_a(\frac{r}{2})+\phi_b(-\frac{r}{2}))} \rangle_{\text{eff.}}$.
In $d=3$ and at   low temperature we finally obtain
\begin{eqnarray}
   C^{RS}_{\text{tripl.}}(r)
   \sim  \frac{2 \pi  }{c^2 a r } e^{-  \frac{\tilde{B}}{2}}.
\end{eqnarray}
It is interesting to evaluate, at this stage, the relative weight
of the two antisymmetric terms in a satellite peak.
We introduce
the Fukuyama-Lee length (or Larkin-Ovchinikov length)
\cite{fukuyama_pinning,larkin_ovchinnikov_pinning}  $R_a= (c^2 / D)^{1/4-d}$
(for $d=3$  $R_a =c^2  / D$)   such that $\phi$ varies on scale given
 by the length $R_a$.  The
ratio of the two intensity peaks is:
\begin{eqnarray}
 \frac{I_{\text{tripl.}}}{I_{\text{a}}} = - \frac{\Delta f}{\pi  \overline{f}}
 \sqrt{N_I}  \sqrt{\frac{a}{R_a }} .
\label{stima}
\end{eqnarray}
For weak disorder  $R_a \gg a$ it follows that $ I_{\text{a}}  \gg
I_{\text{tripl.}}$ and we thus need only to consider $I_{\text{a}}$  and
$I_{\text{d}}$.

Since for $d=3$ the RS solution is unstable, to obtain the correct
physics one has to look for the RSB method.
Within this scheme \cite{giamarchi_vortex_long},
 the off diagonal elements of $G_{ab}(q)$ are
parameterized  by $G(q,v)$ where $0<v<1$ and the solution is
characterized by a variational breakpoint $v_c$.  The form of the
symmetric part is given in \cite{giamarchi_vortex_long}:
\begin{eqnarray} \label{displRSB}
 C_{\text{d}}(r)\sim e^{-\frac{\phi^{2}_T}{2}}
 \left(\frac{l}{r}\right)^{\eta}
\end{eqnarray}
where $\phi^{2}_T \simeq \frac{2 T}{\pi c a}$  measures the
strength of thermal fluctuations, and $\eta \sim 1$ is the Bragg
glass exponent in $d=3$. At low temperature one has  $l \sim R_a$.
The algebraic behavior of (\ref{displRSB}) is controlled by small
$v$ ($v<v_c$). Values of $v$ above the breaking point ($v>v_c$)
give the small distance contribution. Finally
one finds $v_c=\frac{1}{8} \phi^{2}_T \frac{a}{l}$.

To fully characterize the spectrum  it still remains
to evaluate $\chi(r)$ in the RSB scenario :
\begin{eqnarray}
\label{RSB}
\chi(r)  = \frac{1}{T}[1 - \int_0^1 dv e^{-T  \int_q
    (\tilde{G}(q) - G(q,v)) \cos qr}].
\end{eqnarray}
Restricting to the case $d=3$, we write:
\begin{equation}\label{RSB3d}
 \tilde{G}(q)-G(q,v) =\frac{1}{c}  \left[ \frac{1}{q^2+l^{-2}} +
 \frac{2}{l^2}\int_{v/v_c}^1 dt \frac{1}{\left(q^2 +(\frac{t}{l})^2 \right)^2}\right].
\end{equation}
By integrating (\ref{RSB3d}) over  q and with some manipulations,
(\ref{RSB}) becomes:
\begin{eqnarray}
 \chi(r) &=&    \frac{v_c}{T} \left[ 1- \int_0^{1} dz   \exp \left(-8\pi^3  \int^{1}_z \frac{dt}{t}
 e^{-rt/l} \right)    \right]
\end{eqnarray}

\begin{figure}
\centerline{\includegraphics[width=\figwidth]{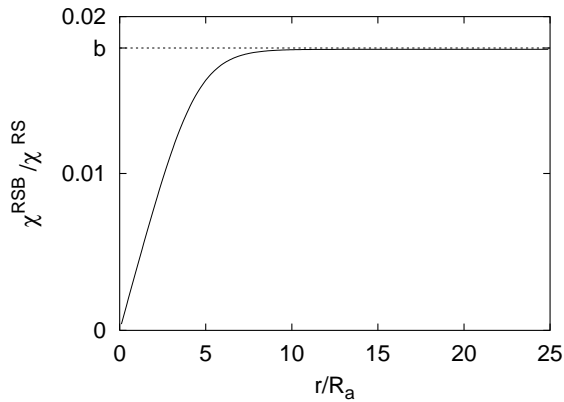} }
\caption{Ratio between the RSB and RS solutions for $\chi(r)$. At
large distance this ratio tends towards a constant value $b$,
where $b \sim 0.018$.
 This means that  the RSB solution affects $\chi(r)$ only by a multiplicative factor.}
\label{integral}
\end{figure}

The low temperature  behavior ($l \sim R_a$) of
this term is sketched in Fig.~\ref{integral}.
As for the Replica Symmetry case, we have  $C_{\text{a}}(r) \propto \frac{1}{r}
e^{-  \frac{ \tilde{B}(r)}{2}}$. We can now compare the two terms:
\begin{eqnarray}
\label{stima2}
I_{\text{d}}(K+Q) & = & \overline{f}^2  K^2  {u_0}^2 \int_r (\frac{R_a}{r})^{\eta}  \\
 I_{\text{a}}(K+Q) & = & - 2\pi^2 \overline{f} {\Delta f} \sqrt{N_I}  \sqrt{\frac{a}{R_a }}
 K u_0 \int_r
  (\frac{R_a}{r})^{\eta} \frac{ba}{r}\nonumber .
\end{eqnarray}
After executing the $d$-dimensional Fourier Transforms, we
conclude that both terms are divergent: in particular,
$I_{\text{d}} \propto \frac{1}{q^{d-\eta}}$ and $I_{\text{a}}
\propto \frac{1}{q^{d-\eta-1}}$.
This effect, shown in Fig.~\ref{model}, is a clear sign of a
quasi-long range positional ordered phase. We have found that the
peak at $K+Q$ is smaller than the $K-Q$ one, as the  potential
between the impurity and CDW is repulsive (we would  have  the
opposite asymmetry in case of an attractive potential). We observe
that for an ideal infinite resolution experiment, the symmetric
term would be dominant, since  $C_{\text{d}} (r)$ decays to zero
less rapidly than $C_{\text{a}}(r)$. However, if the divergence in
(\ref{stima2}) is cut by the finite resolution of the experiment
both terms should be taken into account because $I_d$ is quadratic
in the small parameter $K u_0$ whereas $I_a$ is only linear.

The powerlaw lineshape is obtained for a short range elasticity.
If the Coulomb interaction is unscreened, as might be the case in
fully gapped systems such as the blue bronzes, the dispersion of
$c_1$ should be kept in (\ref{eq:anisotropy}). In that case
$c_1(q) \sim q_x^2/q^2$, which leads to peaks diverging even
faster than (\ref{stima2}) \cite{rosso_cdw_long}.

On the experimental side few detailed diffraction spectra are
available at the moment. One case is doped blue bronzes where the
lineshape corresponding to the CDW has been obtained after
substraction of a Friedel oscillation contribution
\cite{rouzieres_structue_cdw}. The observed asymmetry of the peaks
would be compatible with both strong and weak pinning. However
given the short correlation length extracted from the data, this
particular experiment is most likely still in the strong pinning
regime. It would thus be highly desirable to have more detailed
analysis of the lineshapes either in this compound, for different
impurity concentrations, or in less disordered systems, where one
can expect a Bragg glass behavior.



We thank J.-P.~Poujet and S.~Ravy  for stimulating discussions.

\end{document}